\documentclass[traditabstract]{aa}
\usepackage{graphicx}
\usepackage{natbib}
\usepackage{txfonts}
\usepackage{ulem}
      \def\new#1 {{\bf #1 }}
      \def\cut#1 {\sout{#1} }

\def\sol {$_{\odot}$}

\def\percc {$\mathrm{cm^{-3}}$} 
\def\cmsq  {$\hbox{{\rm cm}}^{-2}$}    

\def\AMM {\hbox{${\rm NH}_{3}$}} 

\def\folio{\ifnum\pageno=1\nopagenumbers\else\number\pageno\fi}

\def\lax    {\ifmmode{_<\atop^{\sim}}\else{${_<\atop^{\sim}}$}\fi}
\def\gax    {\ifmmode{_>\atop^{\sim}}\else{${_>\atop^{\sim}}$}\fi}
\newbox\grsign      \setbox\grsign=\hbox{$>$} 
\newdimen\grdimen   \grdimen=\ht\grsign
\newbox\simgreatbox \setbox\simgreatbox=\hbox{\raise.5ex\hbox{$>$}\llap
                        {\lower.5ex\hbox{$\sim$}}}\ht1=\grdimen\dp1=0pt
\newbox\simlessbox  \setbox\simlessbox =\hbox{\raise.5ex\hbox{$<$}\llap
                        {\lower.5ex\hbox{$\sim$}}}\ht2=\grdimen\dp2=0pt

\newbox\grsign \setbox\grsign=\hbox{$>$} \newdimen\grdimen \grdimen=\ht\grsign
\newbox\laxbox \newbox\gaxbox
\setbox\gaxbox=\hbox{\raise.5ex\hbox{$>$}\llap
     {\lower.5ex\hbox{$\sim$}}}\ht1=\grdimen\dp1=0pt
\setbox\laxbox=\hbox{\raise.5ex\hbox{$<$}\llap
     {\lower.5ex\hbox{$\sim$}}}\ht2=\grdimen\dp2=0pt
\def\gax{\mathrel{\copy\gaxbox}}
\def\lax{\mathrel{\copy\laxbox}}
\def\boxit#1    {\vbox{\hrule\hbox{\vrule\kern3pt
                  \vbox{\kern3pt#1\kern3pt}\kern3pt\vrule}\hrule}}
\def\h      {\ifmmode{^{\rm h}}\else{$^{\rm h}$}\fi}
\def\m      {\ifmmode{^{\rm m}}\else{$^{\rm m}$}\fi}
\def\s      {\ifmmode{^{\rm s}}\else{$^{\rm s}$}\fi}
\def\decas    {\ifmmode{{\rlap.}{''}}\else{${\rlap.}{''}$}\fi}
\def\mum     {\ifmmode{\mu{\rm m}}\else{$\mu{\rm m}$}\fi}
\def\s      {\ifmmode{^{\rm s}}\else{$^{\rm s}$}\fi}
\def\deg      {\ifmmode{^{\circ}}\else{$^{\circ}$}\fi}
\def\as     {\ifmmode {\rlap.}$\,$''$\,$\! \else ${\rlap.}$\,$''$\,$\!$\fi}
\def\decsec  {\ifmmode {\rlap.}$\,$^{s}$\,$\! \else ${\rlap.}$\,$^{s}$\,$\!$\fi}\def\decs  {\ifmmode {\rlap.}$\,$^{s}$\,$\! \else ${\rlap.}$\,$^{s}$\,$\!$\fi}

\def\kms    {\ifmmode{{\rm km~s}^{-1}}\else{km~s$^{-1}$}\fi}

\def\Mspy   {\ifmmode {M_{\odot} {\rm yr}^{-1}} \else $M_{\odot}$~yr$^{-1}$\fi}
\def\Mdot   {\ifmmode {\dot M} \else $\dot M$\fi}
\def\mhd    {\ifmmode {n_{{\rm H}_2}} \else $n_{{\rm H}_2}$\fi}
\def\mhcd   {\ifmmode {N_{{\rm H}_2}} \else $N_{{\rm H}_2}$\fi}

\def\El      {\ifmmode{E_{\ell}}\else{$E_{\ell}$}\fi}
\def\beam    {\ifmmode{\theta_{\rm B}}\else{$\theta_{\rm B}$}\fi}
\def\mjyb   {\ifmmode {{\rm mJy~beam}^{-1}} \else{mJy~beam$^{-1}$}\fi}
\def\mujyb   {\ifmmode {\mu{\rm Jy~beam}^{-1}} \else{$\mu$Jy~beam$^{-1}$}\fi}
\def\Trot   {\ifmmode{T_{\rm rot}}\else$T_{\rm rot}$\fi}    
    
\def\Teff   {\ifmmode{T_{\rm eff}}\else$T_{\rm eff}$\fi}

\def\ITRS   {\ifmmode{\smallint {\rm T}_{R}^{*}dv}\else{$\smallint 
{\rm T}_{R}^{*}dv$}\fi}
\def\ITRS   {\ifmmode{\smallint {\rm T}_{R}^{*}dv}\else{$\smallint 
{\rm T}_{R}^{*}dv$}\fi}
\def\ITAS   {\ifmmode{\smallint {\rm T}_{A}^{*}dv}\else{$\smallint 
{\rm T}_{A}^{*}dv$}\fi}

\def\lefttitle#1  {\noindent \hangindent=18.0pt \hangafter=1 {#1} \par}
\def\vol#1  {{\bf {#1}{\rm,}\ }}
\font\tenssb=cmssbx10
\font\tenbf=cmbx10
\font\sevenbf=cmbx8
\font\fivebf=cmbx6
\def\unetdemi    {\smallskipamount=6pt plus2pt minus2pt
                  \medskipamount=12pt plus4pt minus4pt
                  \bigskipamount=24pt plus8pt minus8pt
                  \normalbaselineskip=16pt plus0pt minus0pt
                  \normallineskip=2pt
                  \normallineskiplimit=0pt
                  \jot=6pt
                  {\def\smallskip {\vskip\smallskipamount}}
                  {\def\medskip   {\vskip\medskipamount}}
                  {\def\bigskip   {\vskip\bigskipamount}}
                  {\setbox\strutbox=\hbox{\vrule 
                    height17.0pt depth7.0pt width 0pt}}
                  \parskip 12.0pt
                  \normalbaselines}
\def\smallerspace {\smallskipamount=3pt plus0pt minus0pt
                  \medskipamount=6pt plus0pt minus0pt
                  \bigskipamount=10.5pt plus0pt minus0pt
                  \normalbaselineskip=10.5pt plus0pt minus0pt
                  \normallineskip=1pt
                  \normallineskiplimit=0pt
                  \jot=3pt
                  {\def\smallskip {\vskip\smallskipamount}}
                  {\def\medskip   {\vskip\medskipamount}}
                  {\def\bigskip   {\vskip\bigskipamount}}
                  {\setbox\strutbox=\hbox{\vrule 
                    height8.5pt depth3.5pt width 0pt}}
                  \parskip 0pt
                  \normalbaselines}
\def\memospace    {\smallskipamount=4pt plus1pt minus1pt
                  \medskipamount=6pt plus2pt minus2pt
                  \bigskipamount=14pt plus6pt minus6pt
                  \normalbaselineskip=14pt plus0pt minus0pt
                  \normallineskip=1pt
                  \normallineskiplimit=0pt
                  \jot=4pt
                  {\def\smallskip {\vskip\smallskipamount}}
                  {\def\medskip   {\vskip\medskipamount}}
                  {\def\bigskip   {\vskip\bigskipamount}}
                  {\setbox\strutbox=\hbox{\vrule 
                    height17.0pt depth7.0pt width 0pt}}
                  \parskip 2.0pt
                  \normalbaselines}
\def\memowidespace    {\smallskipamount=5pt plus1pt minus1pt
                  \medskipamount=7.5pt plus2pt minus2pt
                  \bigskipamount=17.5pt plus6pt minus6pt
                  \normalbaselineskip=17.0pt plus0pt minus0pt
                  \normallineskip=1.25pt
                  \normallineskiplimit=0pt
                  \jot=5pt
                  {\def\smallskip {\vskip\smallskipamount}}
                  {\def\medskip   {\vskip\medskipamount}}
                  {\def\bigskip   {\vskip\bigskipamount}}
                  {\setbox\strutbox=\hbox{\vrule 
                    height21.25pt depth8.75pt width 0pt}}
                  \parskip 2.5pt
                  \normalbaselines}
\begin{document}

\title{Terahertz ammonia absorption as a probe of infall in high-mass
  star forming clumps} 

\author{F. Wyrowski \inst{1} \and R. G\"usten \inst{1} \and
  K. M. Menten \inst{1} \and H. Wiesemeyer \inst{1}  \and B. Klein \inst{1} }

\offprints{F. Wyrowski}

\institute{Max-Planck-Institut f\"ur Radioastronomie,
Auf dem H\"ugel 69, D-53121 Bonn, Germany\\
\email{wyrowski, rguesten, kmenten, hwiese, bklein@mpifr-bonn.mpg.de}
}

\date{Received / Accepted}


\authorrunning{Wyrowski et al.}

\abstract {
  Cloud contraction and infall are the fundamental processes of star
  formation. While ``blue-skewed'' line profiles observed in high-mass
  star forming regions are commonly taken as evidence of infall by an
  ever increasing number of studies, their interpretation offers many
  pitfalls. Detecting infall via redshifted absorption in front of
  continuum sources is a much more direct and reliable method but so
  far mostly restricted toward absorption in the centimeter toward
  strong HII regions. Here we present a novel approach by probing
  absorption of rotational ammonia transitions in front of the strong
  dust emission of massive star-forming regions. A carefully selected
  sample of three regions with different stages of evolution is
  selected to study infall through the evolution of massive
  star-forming clumps. Redshifted absorption is detected toward all
  three sources and infall rates between $3-10\times
  10^{-3}$~M\sol\,yr$^{-1}$ are derived.
}


\maketitle

\section{\label{intro}Introduction}

The earliest phases of massive star formation (MSF) are still poorly
understood. We know that massive stars are born in dense clumps within
giant molecular cloud complexes. Ultracompact HII regions (UCHIIRs)
embedded within these clumps represent a key phase in the early lives
of massive stars \citep[see review by][]{hoare2005}. In their
environments, often hot ($>$ 100~K), compact ($<$0.1~pc), and dense
cores are found, some of which are believed to be in a stage prior to
the formation of UCHIIRs \citep{kurtz+2000}. More recently, so-called
high-mass protostellar objects (HMPOs) or massive young stellar
objects (MYSOs) were recognized to likely represent an even earlier
stage of MSF \citep[e.g.][]{beuther+2002a}. Very recently, clumps
within InfraRed Dark Clouds (IRDCs) were found to be promising
candidates for even earlier stages in the formation of massive stars
\citep[see][and references therein]{menten+2005}, sometimes called
pre-protocluster clumps.

One of the important results of studies of low-mass star formation has
been the observation of infall motions \citep[e.g.][]{belloche+2002},
which give direct evidence of accretion. Models of high mass cluster
formation also predict a large scale contraction as the protocluster
evolves and molecular gas is funneled from the outer regions of the
core into the center of the cluster \citep{bonnell+2002,
  bonnell+2004}. But toward high-mass star-forming cores, the
observational evidence of infall is still scarce
\citep[e.g.][]{wu+2003}.

 Most studies rely on the observations of so-called ``blue-skewed''
 line profiles: with a negative excitation gradient in a molecular clump,
 molecular lines with suitable optical depths (e.g. low $J$ rotational
 lines of HCO$^+$) will show a central self-absorption dip whose 
 blue peak is stronger than the red peak. But there are many
 pitfalls in the interpretation of this signature \citep[see
 discussion in][]{evans2002}. Especially at distances (few kpc)
 typical of MSF sites, other kinematical properties such as rotation and
 outflow activity might easily be mistaken for infall. Furthermore,
 abundance variations in the clumps can change the self-absorption
 profile tremendously. Still, this method led to evidence for
 infalling envelopes of a quite a few of hot cores
 \citep{wyrowski+2006}.

 Alternatively, absorption lines toward compact and strong continuum
 sources can be observed. In particular, centimeter ammonia inversion
 lines have been successfully detected in absorption toward bright
 UCHIIRs, e.g. by \citet{sollins+2005} in 
 G10.6-0.4. They derive gas velocities consistent with material
 spiraling toward the central newly-formed stars. Similar
 results have been reported by \citet{zhang+1997} in the W51 massive
 star-forming region (see their Fig. 6), and by
 \citet{beltran+2006} toward G24.78+0.08. These results strongly
 indicate the presence of infall, as predicted by the accretion
 models. But this method is restricted to a fairly late state of MSF
 when an UCHII has already developed and can be used as a background
 source.

 At THz frequencies new opportunities arise now: Their high dust
 column densities make MSF clumps very bright far-infrared (FIR)
 continuum sources, especially in their early stages. Therefore,
 molecular lines with lower excitation than the (background) continuum
 temperature will be in absorption and will probe the kinematics of
 the clump on the line of sight toward the deeply embedded forming
 massive clusters. This signature is then a much more direct probe of
 infall and can be used through {\it all} embedded stages of MSF.

 Of particular interest is the ammonia molecule \AMM\ which has been
 extensively studied, mostly via its centimeter inversion lines, and
 proven to be one of the best-known interstellar ``thermometers''
 \citep{ho+1983,maret+2009}. In the THz range rotational transitions
 of ammonia can now be observed, in particular transitions from
 non-metastable to metastable levels, which will have low excitation
 temperatures and therefore are likely to be observed in
 absorption. In addition, ammonia is known to be one of the molecules
 that is the least likely to freeze out in the earliest, coldest
 stages of molecular clumps \citep[e.g.][]{bergin+langer1997}.
 
 We therefore present here a first small survey of sources in a range
 of evolutionary stages that we observed with the Stratospheric
 Observatory for Infrared Astronomy (SOFIA) in ammonia to measure their
 infall signatures. This observational evidence of accretion in
 the clumps harboring the forming massive (proto) clusters is
 essential for our understanding of the MSF process.

 \begin{table*}[th]
  \caption{Ammonia source sample and observing parameters.}
  \label{t:sample}
  {\begin{tabular}{llllccccc}
  \\
  \hline
  \\
  Source & Stage       & R.A.(J2000)    & Decl.(J2000)  &  L$_{\rm bol}$  &  $d$  & S(870$\mu$m)$^a$ & $T_{\rm sys}$ & $t_{int}$   \\
        &        & ( h m s ) & ( $^\circ$ $^\prime$ $^{\prime\prime}$ )      &  ($L_\odot$)  &  (kpc) & (mJy/bm) & (K) & (min.)  \\
  \\
\hline
W43-MM1    & MIR-quiet core & 18 47 47.0    & $-$01 54 28   & $2.3\,\times\,10^4$$^b$     & 5.5$^b$ &  21.2 & 3700 & 13 \\
G31.41+0.31& hot core  & 18 47 34.3    & $-$01 12 46   & $2.6\,\times\,10^5$$^c$     & 7.9$^c$ &  21.2 & 3500 & 6.8 \\
G34.26+0.15 & UC H{\tiny II}     & 18 53 18.6    &   +01 14 58   & $4.7\,\times\,10^5$$^c$     & 3.7$^c$ &  44.7 & 3500 & 2.5 \\
\hline
\end{tabular}}\begin{flushleft}
$^a$\citet{schuller+2009}, $^b$\citet{motte+2003}, $^c$\citet{rolffs+2011}.
\end{flushleft}

\end{table*}

\section{\label{obs}Observations and data reduction}

We used the GREAT\footnote{GREAT is a development by the MPI f\"ur
  Radioastronomie and the KOSMA/Universit\"at zu K\"oln, in
  cooperation with the MPI f\"ur Sonnensystemforschung and the DLR
  Institut f\"ur Planetenforschung.} instrument \citep[][this
volume]{heyminck+2012} onboard SOFIA to observe the NH$_3$ $3_{2+} -
2_{2-}$ line at 1810.379971~GHz \citep[taken from the JPL line
catalog,][]{pickett+1998}.  This line has the same lower level as the
well-studied centimeter inversion line at 23.7~GHz of the
($J,K$)=(2,2) level, hence we know that this level is sufficiently
populated to give rise to strong absorption signatures.

To reach this frequency, a slight technical change was necessary: the
modularity of GREAT allows the implementation of new technical
opportunities between flights. A spare local oscillator (LO) source
for the high-frequency channel allowed tuning to slightly lower than
advertised frequencies (1810 instead of 1820~GHz with the nominal
LO). With this setup, the \AMM\ line became accessible, but only if
tuned into the lower sideband of the receiver.
GREAT
operates in double sideband mode and parts of the
image sideband then cover a region of very low transmission, which led to
uncertainties in the determination of the (weak) continuum
level.  The lower frequency channel was tuned to the CO (13--12) line
at 1496.922909~GHz and was in turn used, at much better transmission,
to determine the THz continuum flux of the sources.

The beam sizes at the observing frequencies are 20 and 16 arcsec,
respectively.  The three sources given in Table~\ref{t:sample} were
observed on 2011 July 20 at an altitude ranging from 39000 to 43000 feet.

The system temperatures during the observations and integration times
are given in Table~\ref{t:sample}. As backends, MPIfR fast Fourier
transform spectrometers \citep{Klein+2012} were used to
cover a bandwidth of 1.5~GHz for each channel with a velocity
resolution of 0.03~km/s, which was later lowered to 1~km/s to increase
the signal-to-noise ratio in the spectra. The wobbling secondary was
chopped with a throw of $120''$ about the cross elevation axis in a
symmetric mode. The pointing was monitored with the fiber optic
gyroscopes.  Spectra
were calibrated to a $T_{\rm A}^*$ scale and then the conversion to
$T_{\rm MB}$ was made assuming a forward efficiency of 0.95 and beam
efficiencies of 0.54 and 0.51 for the lower and higher frequency
receivers, respectively.

In addition, complementary ground-based spectra of the C$^{17}$O (3--2)
and HNC (4--3) were obtained with the APEX telescope.

Final processing and analysis of the data was then performed using the
CLASS program within the GILDAS software
\footnote{http://www.iram.fr/IRAMFR/GILDAS}.

\section{Results}
\label{sec:results}

\begin{figure}[ht]
\begin{center}
\includegraphics[width=0.45\textwidth,angle=0]{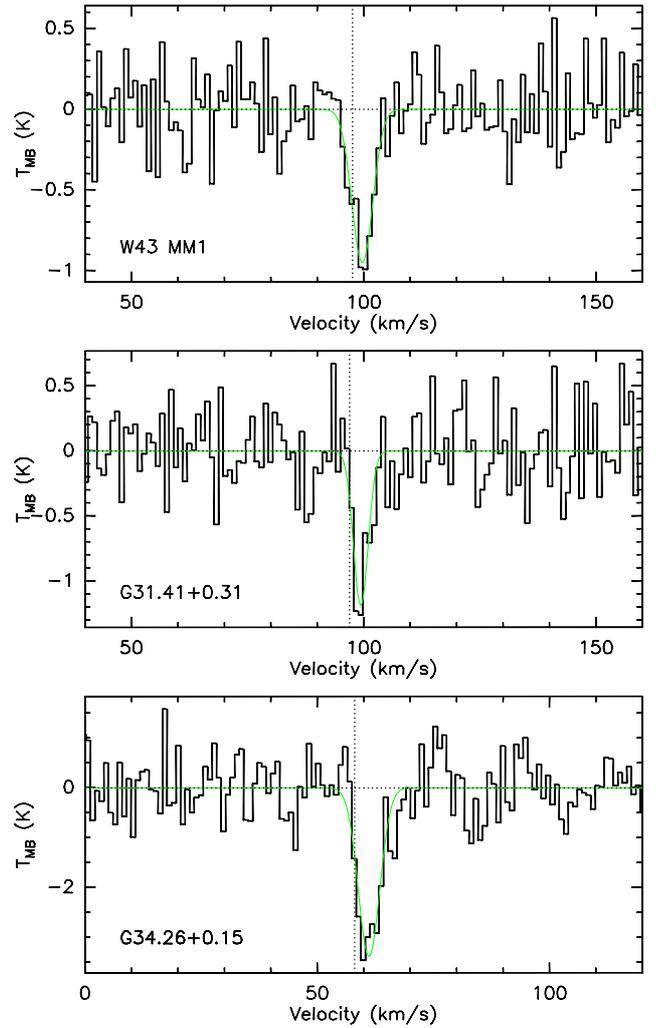}
\caption{\label{fig:spectra} \AMM\ $3_{2+} - 2_{2-}$ spectra of the
  observed sources.  Results of Gaussian fits to the line are overlaid
  in green. The systemic velocities of the sources, determined using
  C$^{17}$O (3--2), are shown with dotted lines.}
\end{center}
\end{figure}

The averaged and baseline-subtracted spectra are shown in
Fig.~\ref{fig:spectra}. Ammonia absorption lines were detected in all
three sources. The line parameters from Gaussian fits are given in
Table~\ref{t:linepar}. Compared with C$^{17}$O (3--2) line velocities,
which as optically thin emission lines probe the systemic velocities
of the sources with a comparable resolution of $18"$, all ammonia
absorption lines are redshifted. This is consistent with blue-skewed
emission profiles observed in HNC (4--3) toward these clumps with the
APEX telescope.  The fact that we observe absorption lines is not
surprising: compared to the microwave inversion lines, the FIR
rotational lines have Einstein A values about five magnitudes higher,
hence they will decay rapidly from the non-metastable (J$>$K) into the
metastable (J=K) levels, which in turn only decay slowly via
collisions \citep{ho+1983}. Thus, the metastable levels are
``overpopulated'' compared with the non-metastable ones.

Because we were unable to measure the 1.8~THz continuum reliably owing
to the double sideband reception and the low atmospheric transmission
in part of the image band, the 1.5~THz continuum was used to estimate
the continuum level at the \AMM\ frequency. Table~\ref{t:contpar}
gives the 1.5~THz continuum level, which was divided by 2 to correct
for the single sideband calibration that by default is applied to the
lines. The modeling of the clumps described in the next section
suggests that the real 1.8~THz continuum might be up to 20\% higher.
The optical depths of the absorption can then be estimated from the
line-to-continuum ratio as $\tau = -\log(1+T_{\rm MB}/T_{\rm cont})$
using the temperatures from Table~\ref{t:linepar} and
\ref{t:contpar}. Optical depths on the order of unity are found meaning that
the filling factor of the absorbing material in front of the dust
continuum must be very high.

The total column density
of the absorbing \AMM\ can then be computed from \citep[e.g.][]{comito+2003},

\begin{equation}
     N{_{\rm 2_2^-}} = 
     \frac{8\pi\nu^3}{A_{ul}c^3}\,\frac{g_l}{g_u}\,\tau\Delta v.
\end{equation} 

This equation is valid for excitation temperatures much lower than
the continuum temperature and also much lower than the
  'transition temperature' $T_{\rm ul}$ ($h\nu/k=86$~K for this \AMM\
  transition), which likely is the case given the rapid radiative
decay of the non-metastable upper level.  The transition probability of 
{$A_{ul}=3.56\times 10^{-2}$~s$^{-1}$} leads to a critical density for the
  transition of $10^9$~\percc. $g_l$ and $g_u$ are the statistical
weights of the lower and upper levels, respectively. Both inversion
levels have almost the same population for densities above
$10^4$~\percc, hence the total column density in the (2,2) levels is
twice as high. This leads to column densities of $1-2\times
10^{14}$~\cmsq\ as given in Table~\ref{t:contpar}.

\begin{table}
  \caption{Line parameters from Gaussian fits to the \AMM\ lines. Nominal fit errors are given in parenthesis. In addition, the velocity of C$^{17}$O (3--2) lines observed with the APEX telescope are given. 
  }
\label{t:linepar} 
\begin{center}
 \begin{tabular}{lcccc}
 \hline\hline 
Source & $T_{\rm MB}$  & $\Delta v$ & $v^{\rm NH_3}_{\rm LSR}$ & $v^{\rm C^{17}O}_{\rm LSR}$ \\
       & (K) & (\kms)    & (\kms) & (\kms)    \\
\hline 
W43-MM1       &  --0.96 (0.22)    & 5.3 (0.8) &  99.7 (0.4)  & 97.65 (0.06) \\
G31.41+0.31   &  --1.18 (0.29)    & 3.7 (0.8) &  99.4 (0.4)  & 97.02 (0.04) \\
G34.26+0.15   &  --3.38 (0.56)    & 5.5 (0.6) &  61.2 (0.3)  & 58.12 (0.03) \\
\hline
 \end{tabular}
 \end{center}
\end{table}

\begin{table}
  \caption{Continuum temperatures at 1.5~THz, the \AMM\ optical depths
           and the column densities in the (2,2) levels. 
  }
\label{t:contpar} 
\begin{center}
 \begin{tabular}{lccc}
 \hline\hline 
Source & $T_{\rm cont}$  & $\tau$ & $N(2,2)$ \\
       & (K) &        & ($10^{14}$\cmsq)   \\
\hline 

W43-MM1       &  1.5    & 1.0 &  1.2 \\
G31.41+0.31   &  1.3    & 2.4 &  2.0 \\
G34.26+0.15   &  5.5    & 1.0 &  1.2 \\
\hline
 \end{tabular}
 \end{center}
\end{table}

\begin{figure}[ht]
\begin{center}
\includegraphics[height=0.48\textwidth,angle=-90]{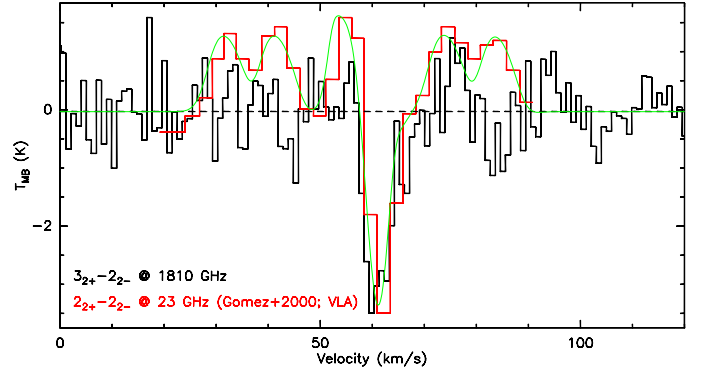}
\caption{\label{fig:vla} G34.26+0.15 SOFIA \AMM\ spectrum compared
  with the VLA \AMM\ (2,2) spectrum (shown in red) taken from
  \citet{gomez+2000}, which was integrated over the region that shows
  absorption. A two-component hyperfine fit to the (2,2) spectrum is
  shown in green.}
\end{center}
\end{figure}

In Figure~\ref{fig:vla} we show a comparison of the (2,2) inversion
line seen with the VLA \citep{keto+1987,gomez+2000} with the SOFIA
results. This inversion line shares the same lower level with the line
observed by SOFIA. The velocity and width of the two absorption lines
agree very well. Interestingly, the VLA spectrum shows the (2,2)
line's outer hyperfine structure components in emission. This can be
understood when the absorbing continuum is only filling the beam
partly and the absorption spectrum is blended with an additional high
optical depth emission component from the hot core that is situated
offset from the continuum.

\section{Analysis}
\label{s:analysis}

To model the observed absorption lines, we used the physical
structures determined by \citet{rolffs+2011}: they fitted the continuum
structure of MSF clumps as observed in the ATLASGAL survey
\citep{schuller+2009} with density power laws $n\sim r^{-1.5}$
  for our observed sources) and determined temperature structures
dictated by the inner heating sources. The velocity structure was
assumed to be a fixed fraction of the free-fall velocities and
adjusted using RATRAN \citep{hogerheijde+2000} to fit high-density
molecular probes observed with the APEX telescope. We compared the
predicted 1.5~THz continuum of the models with the observed
ones. G31.41 was fitted well with dust of \citet{ossenkopf+1994}
  using their models with coagulation at $n=10^5$~cm$^{-3}$ but no ice
  mantles.  For G34.26 we adjusted the grain properties to dust with
thin mantles that have a higher spectral index to reproduce the
high observed continuum temperature. W43-MM1 was not modeled by
\citet{rolffs+2011}. For this source we scaled the G31.41 model to
match the observed 1.5~THz continuum flux.

\begin{figure}[ht]
\begin{center}
\includegraphics[height=0.34\textwidth,angle=-90]{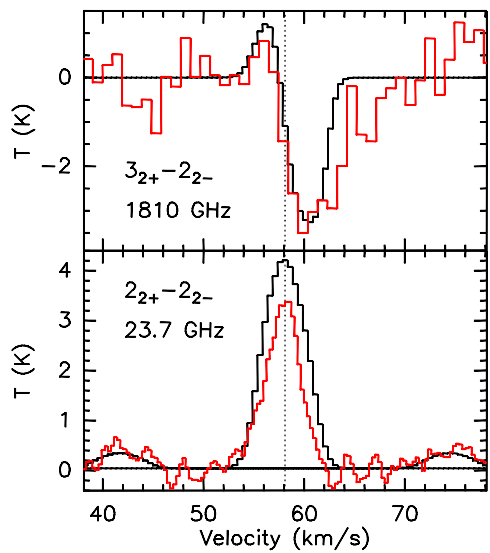}
\caption{\label{fig:g34-mod} G34.26+0.15 \AMM\ SOFIA and Effelsberg
  \citep[taken from][]{churchwell+1990} spectra in red compared with the
  RATRAN model discussed in the text (black profiles). The systemic
  velocity from C$^{17}$O is indicated by a dotted line.}
\end{center}
\end{figure}

In the next step we adjusted the ammonia abundance ($n({\rm
  H_2})/n({\rm NH_3})$) and the velocity to fit the SOFIA \AMM\
absorption lines with the RATRAN code. While in principle the clumps
could have an increased hot core \AMM\ abundance in the inner
$T>100$~K region, this hot core component to the spectrum is
negligible because of its small filling factor. To fit the velocity, we
adjusted the turbulence widths of the gas and the free-fall
  fraction $f$ in $v=-f\times\sqrt{2GM_{\rm in}/r}$, where $M_{\rm
    in}$ is the mass inside $r$ \citep[cf.][]{rolffs+2011}. An
example fit is shown in Fig.~\ref{fig:g34-mod} for G34.26+0.15
together with a fit of the model to the \AMM\ (2,2) inversion line
observed by \citet{churchwell+1990}. In the model, the velocity was
adjusted to 30\% of the free-fall velocity and an ammonia abundance of
$0.8\times 10^{-8}$. The Doppler $b$-parameter was adjusted to
$b=1.5$~\kms. In the SOFIA spectrum, there is even evidence for
additional excess redshifted absorption, not fit by the model, albeit
only with low signal-to-noise ratio. The optically thick main
cm-inversion line is somewhat overpredicted, whereas the outer
hyperfine satellites are reproduced quite well.
High infall velocities in the innermost part of the model do not
contribute to the observed line profile, since the filling factor of
the hot core region is too low.  W43-MM1 and G31.41 could be fitted
with velocity fields with 20\% free-fall contribution.

\section{Discussion and  conclusions}

From the observed and modeled velocity fields the infall rate of the
clumps can be determined. From the RATRAN models, $\dot{M}$ was
determined from the model density and velocity at the radius
corresponding to the SOFIA beam size. Alternatively, Eq.~4 from
\citet{beltran+2006} can be used with the observed column density and
infall velocity.  Both approaches agree and lead to infall rates
between $3-10\times 10^{-3}$~M\sol\,yr$^{-1}$. These rates are high enough
to quench any HII region \citep{walmsley1995} and to overcome the
radiation pressure from the forming (proto) cluster, although we only
probed the infall at the outer envelope. In the inner part accretion
might already be halted.  From the absorption, we also only probed
the infall on the line-of-sight, hence it might not be spherical.

In summary, we present the first velocity-resolved ammonia spectra
from a rotational transition at THz frequency toward MSF clumps.
Absorption of a comparable \AMM\ $3_{2-} - 2_{2+}$ line at 1764~GHz
was also recently observed with Herschel in the low-mass source
IRAS16293-2422 \citep{hily-blant+2010}. In G34.26 the redshifted
absorption agrees with cm absorption observed with the VLA
toward the UCHII component B \citep{heaton+1989}. This component is within
the SOFIA beam but the strong absorption we detected implies that the
absorption fills most of the dust continuum, which peaks toward the hot
molecular core.  G31.41 and W43-MM1 do not have compact cm
continuum that might be used for absorption studies at cm
  wavelengths, hence our observations are first detections of
redshifted absorption in these sources and demonstrate that ammonia
absorption in front of strong dust continuum can be successfully used
to probe infall in a variety of evolutionary stages of MSF clumps.

\acknowledgements{
  Based [in part] on observations made with the NASA/DLR Stratospheric
  Observatory for Infrared Astronomy. SOFIA Science Mission Operations
  are conducted jointly by the Universities Space Research
  Association, Inc., under NASA contract NAS2-97001, and the Deutsches
  SOFIA Institut under DLR contract 50 OK 0901.
}

\bibliographystyle{aa}
\bibliography{great-nh3}

\clearpage

\end{document}